\documentclass[aps,prb,showpacs]{revtex4}

\usepackage{graphicx, amsmath}

\begin{document}

\title{Electric dipole moment oscillations in Aharonov-Bohm quantum rings}

\author{A. M. Alexeev}
\affiliation{School of Physics, University of Exeter, Stocker Road,
Exeter EX4 4QL, United Kingdom}

\author{M. E. Portnoi}
\email[Corresponding author: ]{M.E.Portnoi@exeter.ac.uk}
\affiliation{School of Physics, University of Exeter, Stocker Road,
Exeter EX4 4QL, United Kingdom}
\affiliation{International Institute of Physics, Av. Odilon Gomes de Lima, 1722, Capim Macio, CEP: 59078-400, Natal - RN, Brazil}

\date{1 June 2012}

\begin{abstract}
Magneto-oscillations of the electric dipole moment are predicted and analyzed for a single-electron nanoscale ring
pierced by a magnetic flux (an Aharonov-Bohm ring) and subjected to an electric field in the ring's plane.
These oscillations are accompanied by periodic changes in the selection rules for inter-level optical
transitions in the ring allowing control of polarization properties of the associated terahertz radiation.
\end{abstract}


\maketitle

\section{Introduction}
Progress in epitaxial techniques has resulted in burgeoning developments in the physics of quantum dots, i.e., semiconductor-based `artificial atoms'. More recently a lot of attention has been turned towards non-simply-connected nanostructures, i.e., quantum rings, which have been obtained in various semiconductor systems.\cite{Lorke,Ribeiro,Chen} The fascination with quantum rings is partially caused by a wide variety of purely quantum-mechanical effects, which are observed in ring-like nanostructures (for a review, see Refs.~\onlinecite{review1,review2,review3}). The star among them is the Aharonov-Bohm effect,\cite{Siday,Aharonov-Bohm} in which a charged particle is influenced by a magnetic field away from the particle's trajectory, resulting in magnetic-flux-dependent oscillations of the ring-confined particle energy. The oscillations of the single-particle energy are strongly suppressed by distortion of the ring shape or by applying an in-plane (lateral) electric field, thus reducing the symmetry of the system. \cite{Barticevic,Bruno-Alfonso} However, there are other physical quantities, which might have even more pronounced magneto-oscillations when the symmetry of the ring is reduced. For example, in the presence of a lateral electric field exceeding a particular threshold it is possible to switch the ground state of an exciton in an Aharonov-Bohm ring from being optically active (bright) to optically inactive (dark).\cite{Fischer,Vivaldo}
Another hitherto overlooked phenomenon is the flux-periodic change of an electric dipole moment of a quantum ring, which is the main subject of this work.

In Sec.~\ref{S2}, we discuss the single-electron energy spectrum of an infinitely-narrow Aharonov-Bohm ring subjected to a lateral electric field.
In Sec.~\ref{S3}, we consider magneto-oscillations of the ring's electric dipole moment and study their electric field and temperature dependence. Matrix elements of the dipole moment calculated between different states define the selection rules for optical transitions. For experimentally attainable quantum rings these transitions occur at terahertz (THz) frequencies. In Sec.~\ref{S4}, we discuss optical selection rules and show how the polarization properties of the associated THz radiation can be tuned by external electric and magnetic fields. Section \ref{S5} contains a brief discussion of the potential applications of the predicted effects. Whereas all of the results presented in the main body of the paper are based on numerical diagonalization of large-size matrices, in the Appendix we provide an analytical treatment of several of the lowest eigenstates using $3\times3$ and $2\times2$ matrices, which yields a clear physical picture with only a marginal loss of accuracy.

\section{Energy spectrum of a quantum ring in a lateral electric field}
\label{S2}
The Hamiltonian of an electron confined in an infinitely narrow quantum ring pierced by magnetic flux $\Phi$ depends only on the polar coordinate $\varphi$
\begin{equation}
\label{Hamiltonian}
\widehat{H}_{\Phi}=- \frac {\hbar^{2}} {2 M_{e} R^{2}} \frac {\partial^{2}} {\partial \varphi^{2}} - \frac {i \hbar e} {2 \pi} \frac {\Phi} {M_{e} R^{2}} \frac {\partial} {\partial \varphi} + \frac {e^{2} \Phi^{2}} {8 \pi^{2} M_{e} R^{2}}\, \mbox{,}
\end{equation}
where $M_{e}$ is the electron effective mass and $R$ is the ring radius.

\begin{figure}[t]%
\includegraphics*[width=0.4\textwidth]{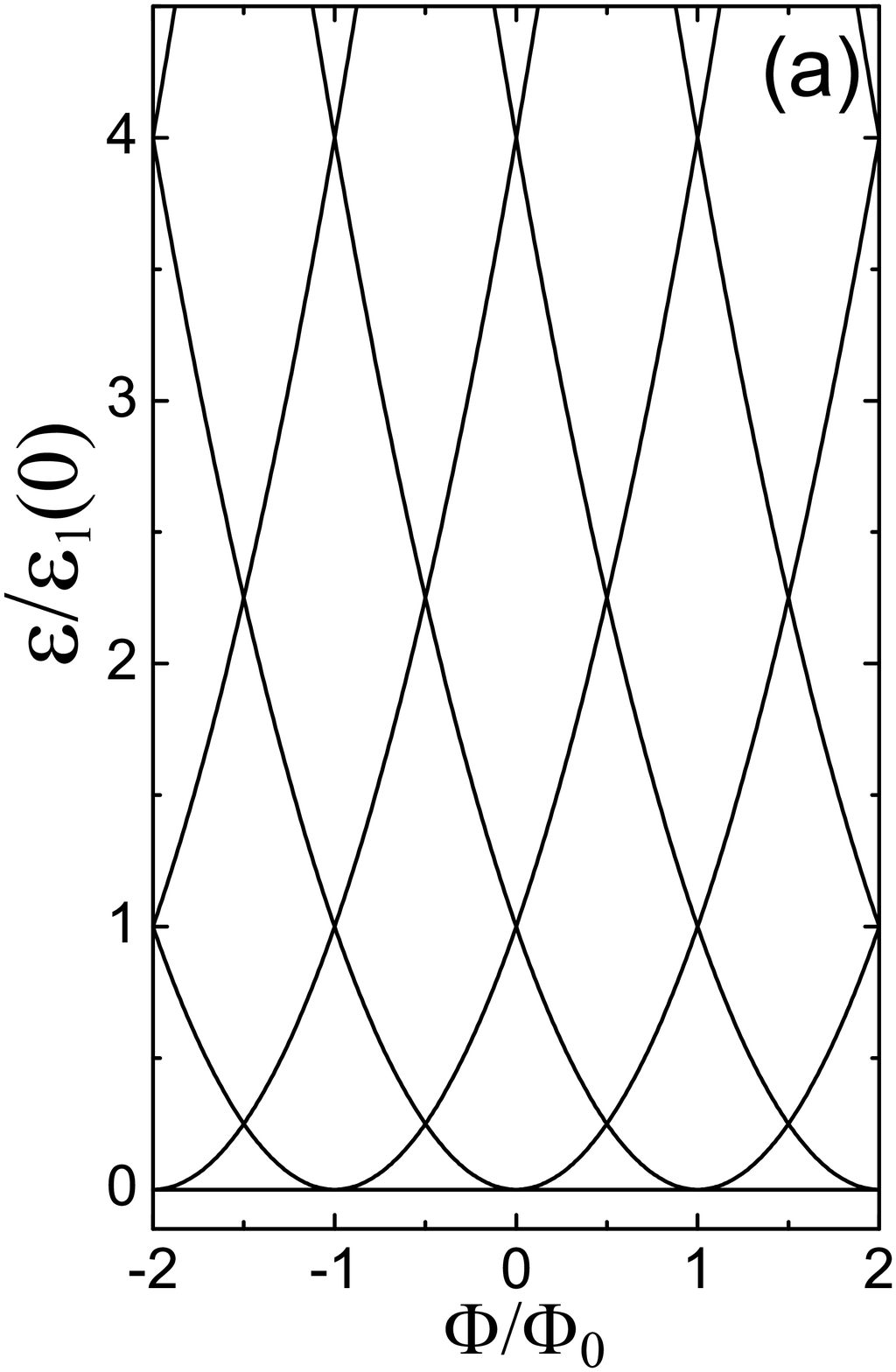}
\includegraphics*[width=0.4\textwidth]{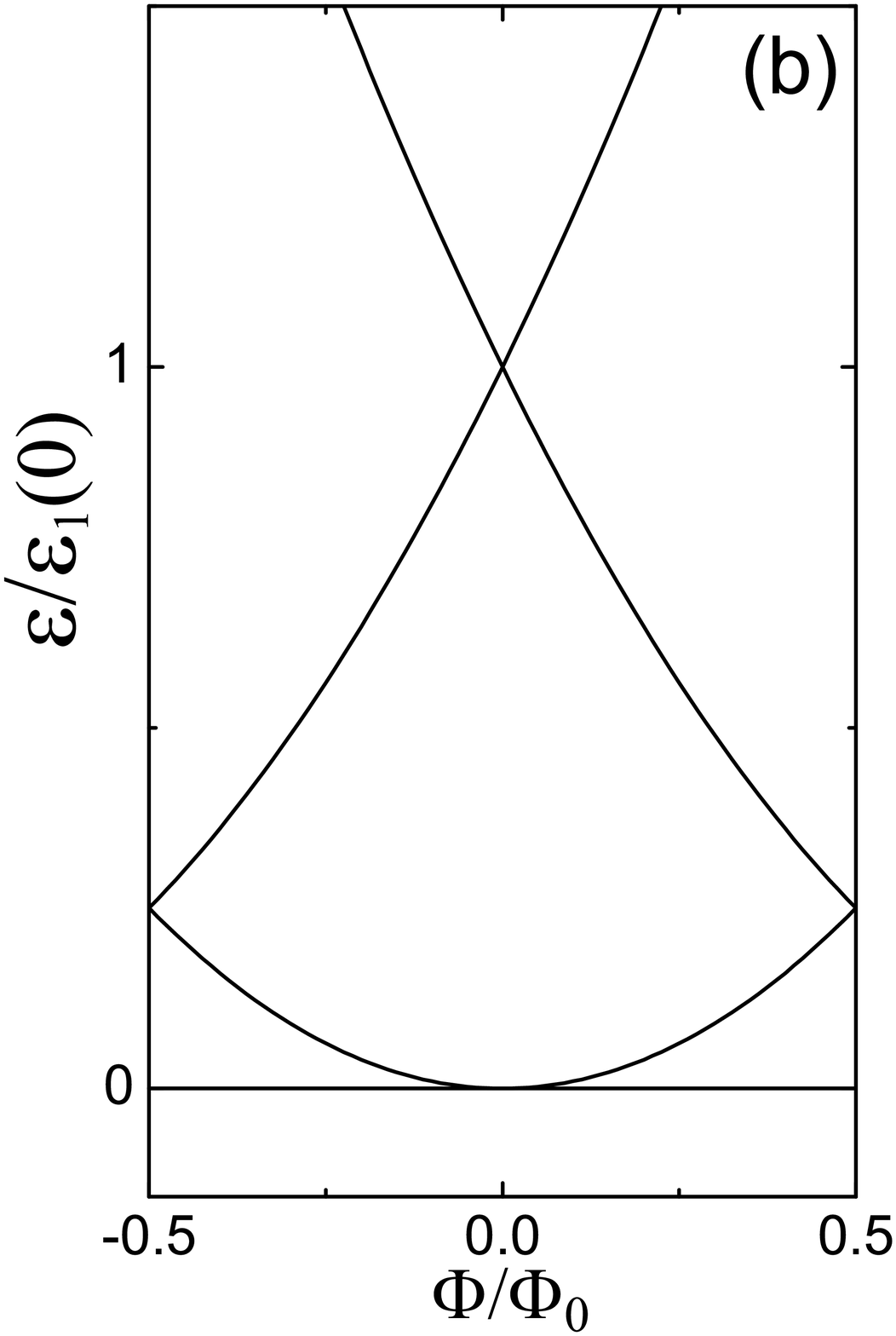}
\caption{%
  (a) The energy spectrum of an infinitely narrow quantum ring pierced by a magnetic flux $\Phi$. Each parabola corresponds to a particular value of the electron angular momentum $m$. The electron energies $\varepsilon$ are plotted versus the number of flux quanta $\Phi / \Phi_{0}$. (b) Expanded view on a smaller energy scale.}
\label{pic:ESwithoutEF}
\end{figure}

The $2\pi$-periodic eigenfunctions of the Hamiltonian defined by Eq.~(\ref{Hamiltonian}) are
\begin{equation}
\label{WFwithoutEF}
\psi_{m} \left ( \varphi \right ) = \frac {e^{i m \varphi}} {\sqrt{2 \pi}}\, \mbox{,}
\end{equation}
and the corresponding eigenvalues are given by
\begin{equation}
\label{EwithoutEF}
\varepsilon_{m}(f) = \frac {\hbar^{2}  \left ( m + f \right ) ^{2}} {2 M_{e} R^{2}}=
\left ( m + f \right )^{2} \varepsilon_{1}(0) \, \mbox{.}
\end{equation}
Here $m= 0, \pm 1, \pm 2 ...$ is the angular momentum quantum number, and $f=\Phi / \Phi_{0} $ is the number of flux quanta piercing the ring ($\Phi_{0}= h / e $). The electron energy spectrum defined by Eq.~(\ref{EwithoutEF}) is plotted in Fig.~\ref{pic:ESwithoutEF}.
It exhibits oscillations in magnetic flux with the period equal to $\Phi_{0}$, known as Aharonov-Bohm oscillations.\cite{Lorke,Aharonov-Bohm} One can see intersections (degeneracy) of the energy levels with different angular momenta, when $\Phi$ is equal to an integer number of $\Phi_{0}/2$. Optical selection rules allow transitions between states with angular momentum quantum numbers different by unity
($\Delta m = \pm 1$). For typical nanoscale rings \cite{Lorke,Ribeiro} the energy scale of the inter-level separation, $\varepsilon_{1}(0)= \hbar^{2} / 2 M_{e} R^{2}$, is in the THz range. When $\Phi$ exceeds $\Phi_{0}/2$ the electron possesses a non-zero angular momentum in the ground state.

Applying an in-plane electric field $\mathbf{E}$ removes the circular symmetry of the system. An additional term corresponding to the electric field appears in the Hamiltonian, which acquires a form
\begin{equation}
\widehat{H}=\widehat{H}_{\Phi} + eER\cos{\varphi} \mbox{.}
\label{HwithE}
\end{equation}
Now the angle $\varphi$ is counted from the direction of the electric field. The field mixes electron states with different angular momentum, which is not a good quantum number anymore.
An eigenfunction of the Hamiltonian (\ref{HwithE}), which maintains the $2\pi$-periodicity in $\varphi$, can be written as a linear combination of the wavefunctions (\ref{WFwithoutEF})
\begin{equation}
\label{WFinpresenceofEF}
\Psi_{n} \left ( \varphi \right ) = \sum \limits_{m} c_{m}^{n} e^{i m \varphi} \mbox{.}
\end{equation}
Substituting the wavefunction (\ref{WFinpresenceofEF}) into the Schr\"{o}dinger equation with the Hamiltonian (\ref{HwithE}), multiplying the resulting expression by $e^{-i m \varphi}$, and integrating with respect to $\varphi$ leads to an infinite system of linear equations for the coefficients $c_{m}^{n}$,
\begin{equation}
\left [ \left ( m+f \right ) ^{2} - \lambda_{n} \right ] c_{m}^{n} + \beta \left ( c_{m+1}^{n} + c_{m-1}^{n} \right ) = 0\,\mbox{,}
\label{Linearsystem}
\end{equation}
where  $\beta = e E R / 2 \varepsilon_{1}(0)$ and $\lambda_{n} = \varepsilon_{n}/ \varepsilon_{1}(0)$, with $\varepsilon_{n}$ being the $n$th eigenvalue of the Hamiltonian (\ref{HwithE}). It is apparent from Eq.~(\ref{Linearsystem}) that all of the properties of the ring are periodic in magnetic flux. Therefore, it is sufficient to consider $0 \le f \le 1/2$, whereas the calculations for other values of $f$ can be performed by shifting $m$ in Eq.~(\ref{Linearsystem}) by an integer number. Interestingly, exactly the same analysis is applicable to a nanohelix subjected to an electric field normal to its axis.\cite{helix1,helix2,helix3} For a helix the role of magnetic flux is played by the electron momentum along the helical line.

\begin{figure}
\includegraphics*[width=0.4\textwidth]{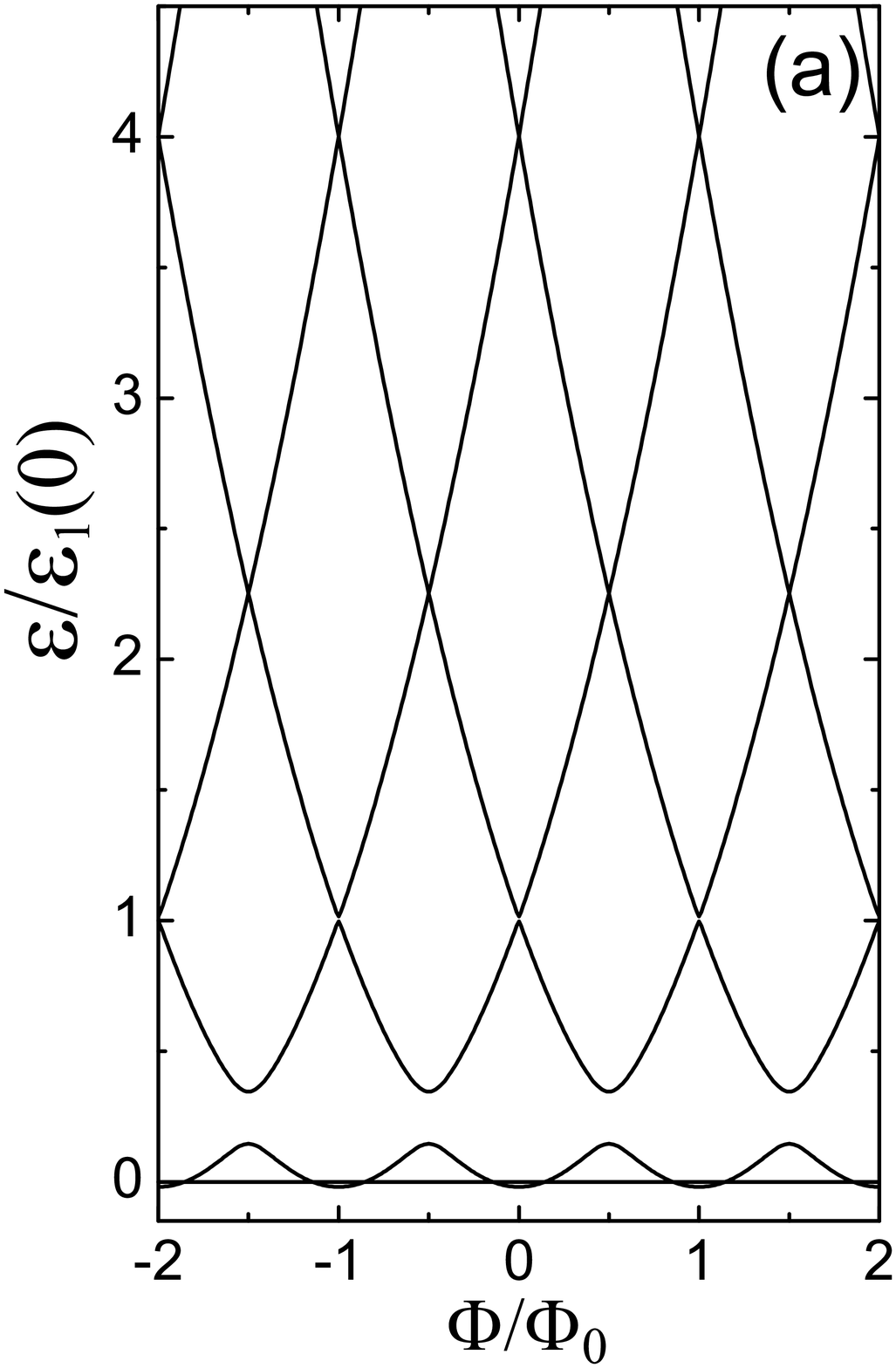}
\includegraphics*[width=0.4\textwidth]{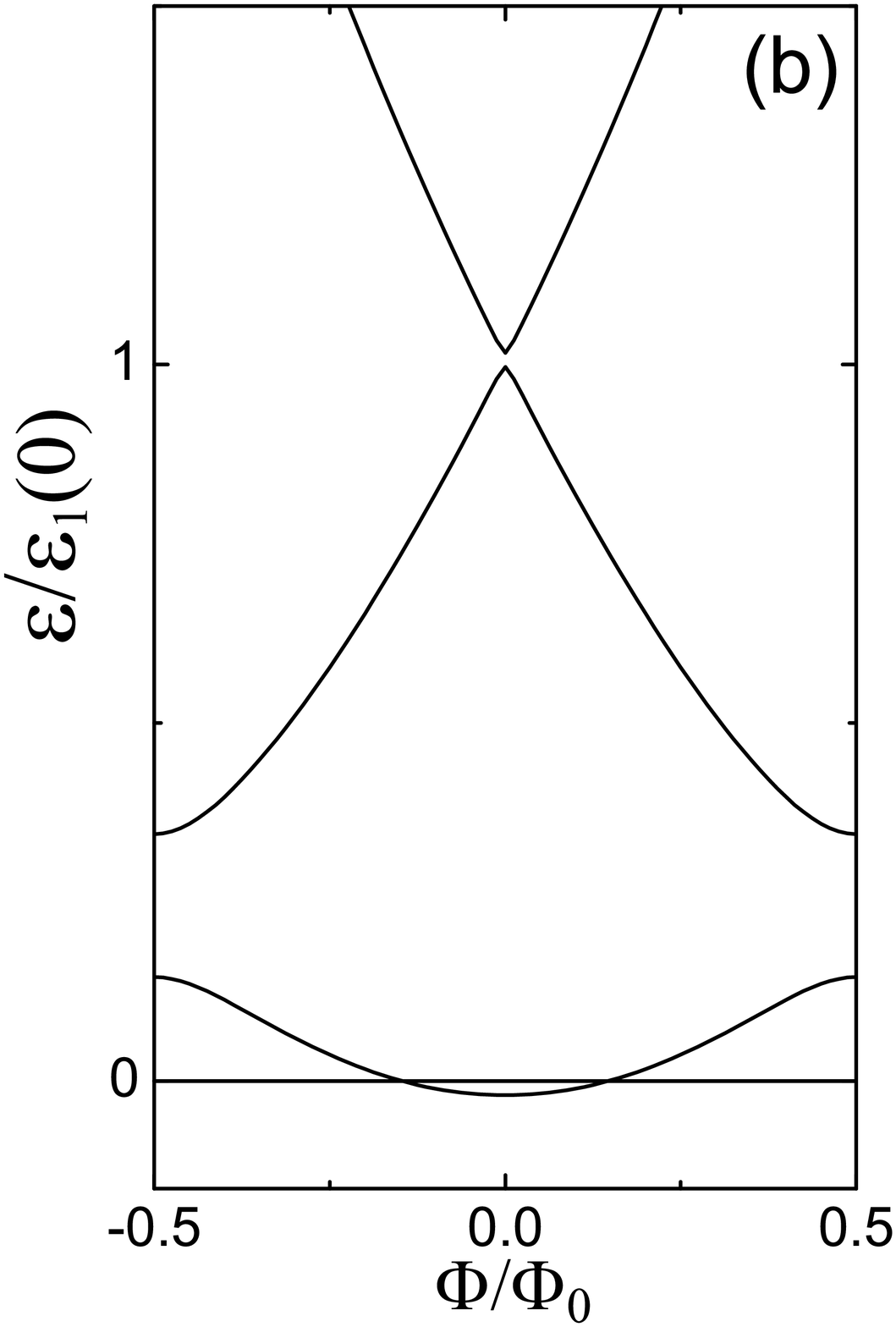}
\caption{%
  (a) The energy spectrum of an infinitely narrow quantum ring of radius $R$ pierced by a magnetic flux $\Phi$ and subjected to an in-plane electric field $E=0.2 \varepsilon_{1}(0)/eR$. The electron energies $\varepsilon$ are plotted versus the number of flux quanta $\Phi / \Phi_{0}$. (b) Expanded view on a smaller energy scale.}
\label{pic:ESinpresenceofEF}
\end{figure}

It should be emphasized that we consider a single-electron problem and are interested only in a few low-energy states. This treatment is relevant to nanoscale-sized semiconductor quantum rings or type-II quantum dots discussed in Refs.~\onlinecite{Lorke,Ribeiro,Chen,review3,Fischer,Vivaldo} and neglects the many-body effects which are known to influence Aharonov-Bohm oscillations in mesoscopic rings.\cite{review1,review2}
The energy levels $\varepsilon_{n}$ as well as the coefficients $c_{m}^{n}$ can be found by cutting off the sum in Eq.~(\ref{WFinpresenceofEF}) at a particular value of $ \left | m \right |$. The results of the numerical diagonalization of the matrix corresponding to the system of linear equations (\ref{Linearsystem}), with a cut-off value of $ \left | m \right |=11$, are plotted in Fig.~\ref{pic:ESinpresenceofEF}. The same cut-off value was chosen in all numerical calculations presented in this paper, since a further increase of the matrix size does not lead to any noticeable change in the results for the three lowest-energy states, which we are interested in.
In small electric fields, $eER \ll \hbar^{2} / 2 M_{e} R^{2}$, a significant change in the ring's energy spectrum occurs only for the ground and two lowest excited states, when $\Phi$ is close to an integer number of $\Phi_{0}/2$ (the points of degeneracy in the absence of the electric field). The most prominent change is associated with the linear in the electric field splitting between the ground and first excited states for half-integer $f$. The less pronounced quadratic in the electric field splitting between the first and second excited states occurs for integer $f$. These splittings can be easily understood with the help of perturbation theory, as there is a non-zero matrix element of $eER\cos{\varphi}$ between the ground and the first excited state, whereas the two excited states are only repelled in the second order via the ground state. As shown in the Appendix, these essential features of the low-energy spectrum are fully captured by considering small-size matrices, which allow an analytical treatment: a $2\times2$ matrix for half-integer $f$ and a $3\times3$ matrix for integer $f$.

As one can see from Fig.~\ref{pic:ESinpresenceofEF}, energy oscillations in the ground state are strongly suppressed even for $e E R=0.2 \hbar^2/2M_{e}R^2$. This suppression is a major source of difficulty in spectroscopic detection of Aharonov-Bohm oscillations. However, as we show in the next two sections, apart from the ground-state energy there are other physical quantities, such as a dipole moment of the ring and polarization properties of the inter-level transitions, which have highly-pronounced magneto-oscillations when the symmetry of the ring is reduced.

\section{Magneto-oscillations of the electric dipole moment}%
\label{S3}
In this section we consider Aharonov-Bohm oscillations of the quantum ring's electric dipole moment.
If an electron occupies the $n$th state of the neutral single-electron quantum ring with a uniform positive background, or if a positive charge $+e$ is placed at the center of the ring, then the projection of the dipole moment on the direction of the lateral electric field is given by
\begin{equation}
P_n = eR\int |\Psi_n|^2 \, \cos{\varphi} \,  d\varphi.
\label{PnI}
\end{equation}
Substituting the wavefunction (\ref{WFinpresenceofEF}) into Eq.~(\ref{PnI}) yields
\begin{equation}
\label{PnS}
P_{n}= \frac {e R} {2} \sum \limits_{m} c_{m}^{n} \left ( c_{m-1}^{n} + c_{m+1}^{n} \right) \mbox{,}
\end{equation}
where the coefficients $c_{m}^{n}$ can be found from the system of linear equations (\ref{Linearsystem}).

In the absence of an electric field, each of the electron states is characterized by a particular value of angular momentum. The electron charge density is spread uniformly over the ring and there is no net dipole moment. The same result is given by Eq.~(\ref{PnS}) -- all of the products $c^n_m c^n_{m\pm1}$ entering Eq.~(\ref{PnS}) vanish for any value of $n$ resulting in the ring dipole moment being equal to zero. Let us now consider what happens to the ground state's dipole moment in the presence of a weak electric field, $eER \ll \hbar^2/2M_{e}R^2$. For $\Phi=0$, the ground state is a practically pure $m=0$ state with a tiny admixture of $m\neq 0$ wavefunctions. However, the situation changes drastically near the points of degeneracy when the magnetic flux through the ring is equal to any odd integer of $\Phi_0/2$. For a half-integer flux, even an infinitely small field modifies entirely the wave function of the ground state. As shown in the Appendix, when $f=1/2$, the ground-state wave-function angular dependence is well-described by $\sin \left (\varphi/2 \right)$. Thus, the ground-state electron density distribution becomes shifted to one side of the ring, against the applied electric field. Such a shift is energetically favorable and results in the value of the dipole moment being close to $eR$. Simultaneously, the first excited state wave-function angular dependence becomes well-described by $\cos \left (\varphi/2 \right)$. For the excited state, the electron is localized near the opposite side of the ring, resulting in a dipole moment of the same magnitude as for the ground state but with the opposite sign.

\begin{figure}[t]%
\includegraphics*[width=0.4\linewidth]{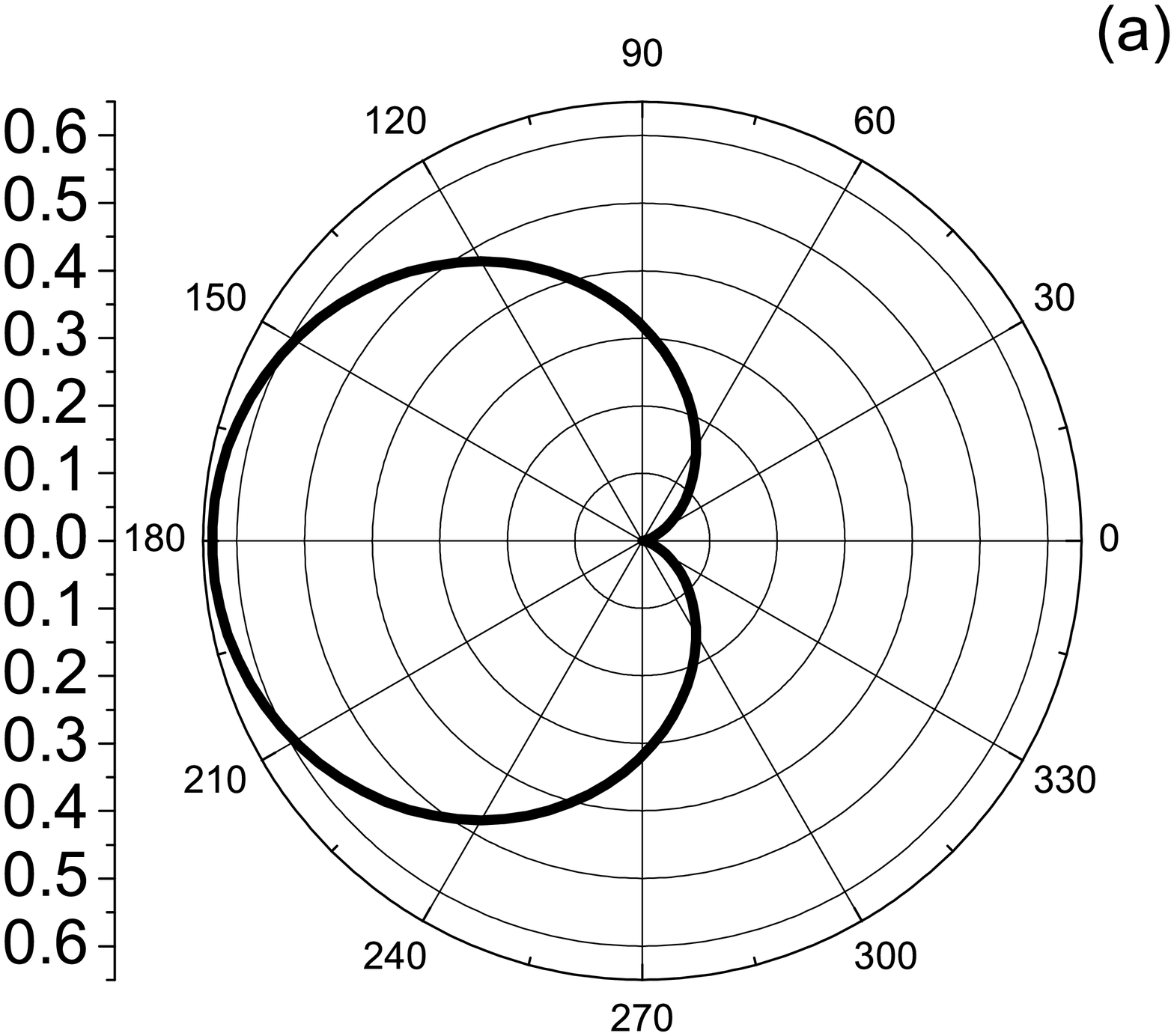}
\includegraphics*[width=0.4\linewidth]{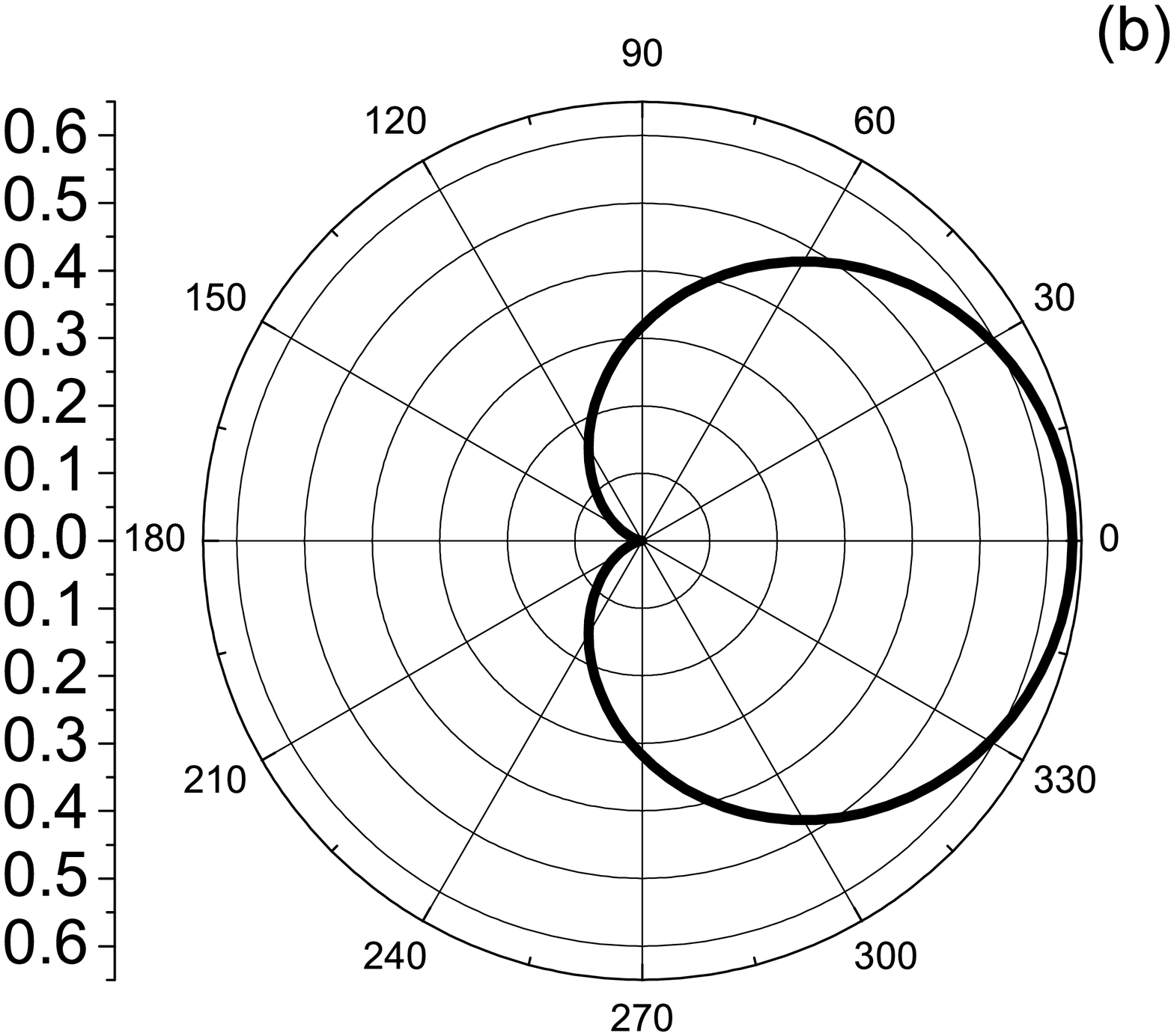}
\caption{%
A polar plot of the electron density distribution in a single-electron quantum ring pierced by a half of the flux quantum and subjected to a weak in-plane electric field, $E\ll\varepsilon_{1}(0)/eR$, applied at the zero angle: (a) for the electron ground state, and (b) for the first excited state.}
\label{pic:WFshape}
\end{figure}

\begin{figure}[t]%
\includegraphics*[width=0.8\linewidth]{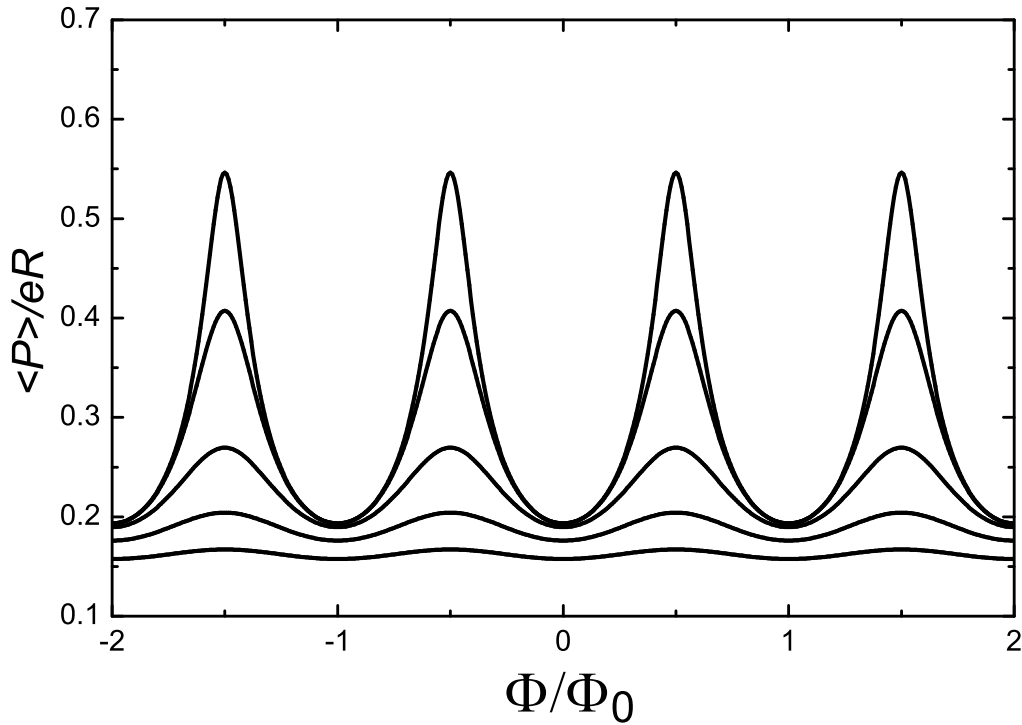}
\caption{%
   Magneto-oscillations of the dipole moment of a ring at various temperatures for $E = 0.2 \varepsilon_{1}(0)/eR$. Different curves correspond to different temperatures in the range from $T = 0.01 \varepsilon_{1}(0) / k_{\mathbf{B}}$ to $T=0.41 \varepsilon_{1}(0) / k_{\mathbf{B}}$ with the increment $0.1 \varepsilon_{1}(0)/k_{\mathbf{B}}$. The upper curve corresponds to $T = 0.01 \varepsilon_{1}(0)/k_{\mathbf{B}}$.}
\label{pic:DMoscillationsT}
\end{figure}

The electron density distributions in the ground and first excited states, when $\Phi=\Phi_0/2$ and the degeneracy is lifted by a weak electric field,
is shown in Fig.~\ref{pic:WFshape}. With changing magnetic flux the ground state density oscillates with a period $\Phi_0$ from an unpolarized to a strongly polarized distribution, resulting in the corresponding dipole moment oscillations. However, the oscillations of the total dipole moment of the ring should be partially compensated if the first excited state, which carries a dipole moment opposite to the ground state's dipole moment for a flux equal to an odd number of $\Phi_{0}/2$, is also occupied due to a finite temperature.
The effect of temperature $T$ can be taken into account by thermal averaging over all states,
\begin{equation}
\label{Paverage}
\left < P \right > = \frac {\sum \limits_{n} P_{n} \exp \left ( - \varepsilon_{n} / k_{\mathbf{B}} T \right)} {\sum \limits_{n} \exp \left ( - \varepsilon_{n} / k_{\mathbf{B}} T \right)} \mbox{.}
\end{equation}
The results of our numerical calculations, using Eq.~(\ref{Paverage}), for several temperature values are shown in Fig.~\ref{pic:DMoscillationsT}.
The dipole moment oscillations, which are well-pronounced for $k_{\mathrm{B}} T \ll e E R$, become suppressed when the temperature increases.

\begin{figure}[t]%
\includegraphics*[width=0.8\linewidth]{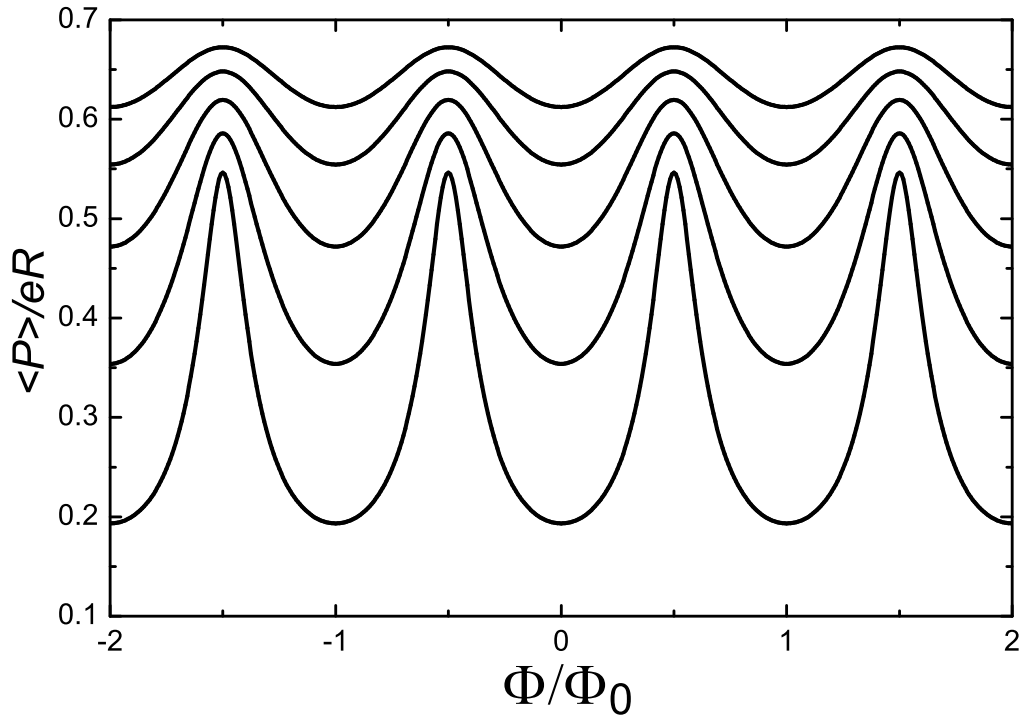}
\caption{%
 Magneto-oscillations of the dipole moment of a ring at various magnitudes of the in-plane electric field for $T= 0.01 \varepsilon_{1} (0)/k_{\mathrm{B}} $. Different curves correspond to different magnitudes of the electric field in the range from $E = 0.2 \varepsilon_{1}(0)/eR$ to $E = 1.0 \varepsilon_{1}(0)/eR$ with the increment $0.2 \varepsilon_{1}(0)/eR$. The upper curve corresponds to $E = 1.0 \varepsilon_{1}(0)/eR$.}
\label{pic:DMoscillationsE}
\end{figure}

In this paper, we consider the limit of weak electric field only. Higher fields, $eER > \hbar^2/2M_{e}R^2$, localize the ground state electron near one side of the ring even in the absence of a magnetic field and the change of magnetic flux through the ring can no longer influence the electron density distribution. For all values of $\Phi$ the ground state wavefunction consists of a mixture of functions with different angular momenta, ensuring that this state is always strongly polarized. The suppression of the dipole moment oscillations with increasing electric field can be seen in Fig.~\ref{pic:DMoscillationsE} where the upper curves, corresponding to higher electric fields and higher dipole moments, exhibit less pronounced oscillations. The energy oscillations for several lowest states are known to be completely suppressed in strong electric fields. \cite{Bruno-Alfonso}

At this point, it would be instructive to discuss conditions needed for an experimental observation of electric dipole moment magneto-oscillations in quantum rings. A typical radius for experimentally attainable rings \cite{Lorke,Ribeiro,Chen} is $R \simeq 20\,$nm. This gives the characteristic energy scale of the inter-level separation $\varepsilon_{1}(0) \simeq 2\,$meV (corresponding to 0.5THz) for an electron of effective mass $M_{e}=0.05m_e$. For a ring with $R=20\,$nm, the magnitude of a magnetic field producing a flux $\Phi=\Phi_{0}$ is $B \simeq 3\,$T.
Therefore, a further decrease of the ring radius would require magnetic fields which are hard to achieve. A typical electric field needed for pronounced dipole moment oscillations is $E = 0.1 \varepsilon_{1}(0) / eR \simeq 10^4\,$V/m, which can be easily created. By far the most difficult condition to be satisfied is the requirement on the temperature regime, $T < eER/ k_{\mathrm{B}}$. For the discussed electric field and ring radius this condition becomes $T<2\,$K. In principle, such temperatures can be achieved in laboratory experiments and magneto-oscillations can be detected, for example, in capacitance measurements. However, for practical device applications, such as quantum-ring-based magnetometery, higher temperatures are desirable. In the next section, we consider a process, that is less sensitive to the temperature-induced occupation of excited states.

\section{Terahertz transitions and optical anisotropy}
\label{S4}
In this section, we study the influence of the in-plane electric field on polarization properties of radiative inter-level transitions in Aharonov-Bohm rings. We restrict our consideration to linearly-polarized radiation and dipole optical transitions only. The case of circular polarization is briefly discussed at the end of the section.

\begin{figure}[h]%
\centering
\includegraphics*[width=0.4\linewidth]{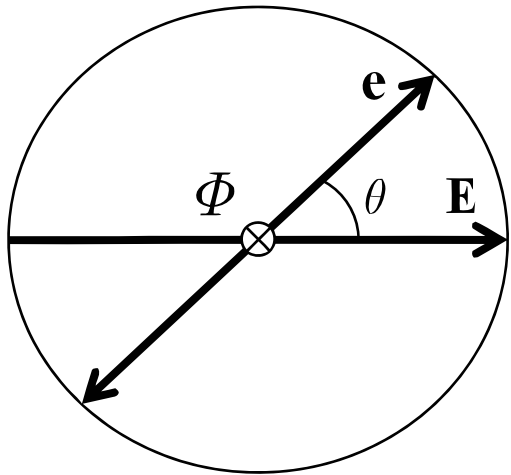}
\caption{%
Relative directions of the external electric field $\mathbf{E}$ and the projection $\mathbf{e}$ of the THz radiation polarization vector onto the quantum ring's plane.}
\label{pic:GoftheP}
\end{figure}

The transition rate $T_{if}$ between the initial ($i$) and final ($f$) electron states is governed by the matrix element
 $P_{if}=\langle f | \mathbf{e} \hat{\mathbf{P}} | i\rangle\,$, where $\hat{\mathbf{P}}$ is the dipole moment operator and $\mathbf{e}$ is the projection of the radiation polarization vector onto the plane of the ring. For our model infinitely-narrow ring
\begin{equation}
\label{MatrixElementI}
P_{if} (\theta) =  e R \int \Psi_{f}^{*} \Psi_{i}  \cos \left ( \theta - \varphi \right ) d \varphi \mbox{,}
\end{equation}
where $\theta$ is the angle between the vector $\mathbf{e}$ and the in-plane electric field $\mathbf{E}$.
The geometry of the problem is shown in Fig.~\ref{pic:GoftheP}.
Substituting the electron wave functions $\Psi_{i}$ and $\Psi_{f}$, given by Eq.~(\ref{WFinpresenceofEF}), into Eq.~(\ref{MatrixElementI}) yields
\begin{equation}
\label{MatrixElementS}
T_{if} \sim P_{if}^{2} \left ( \theta \right ) = {P_{if}^{-}}^{2}+{P_{if}^{+}}^2-2P_{if}^{-} P_{if}^{+} \cos 2 \theta \mbox{,}
\end{equation}
where
\begin{equation}
\label{PifP}
P_{if}^{-}= \frac {e R} {2} \left | \sum \limits_{m} c_{m}^{f} c_{m-1}^{i} \right |
\end{equation}
and
\begin{equation}
\label{PifM}
P_{if}^{+}= \frac {e R} {2} \left | \sum \limits_{m} c_{m}^{f} c_{m+1}^{i} \right | \mbox{.}
\end{equation}
The double angle $2\theta$ entering Eq.~(\ref{MatrixElementS}) ensures that the transition rate does not depend on the sign of $\mathbf{e}$.

\begin{figure}[t]%
\includegraphics*[width=0.8\linewidth]{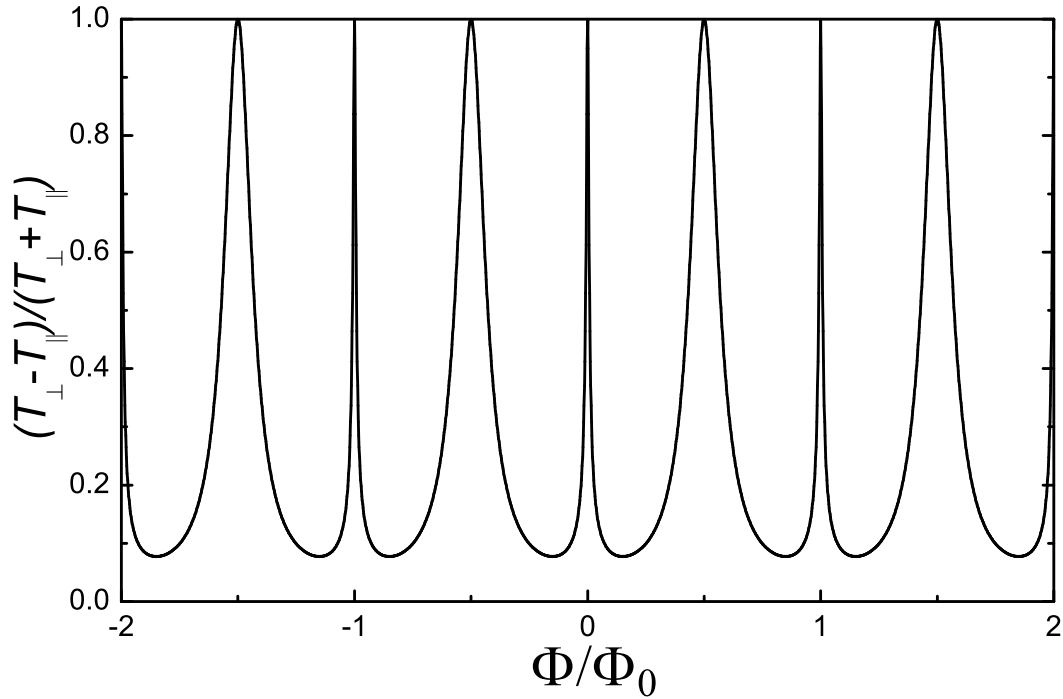}
\caption{%
Magneto-oscillations of the degree of polarization for the transitions between the ground state and the first excited state. Here, $T_{\parallel}$ and $T_{\perp}$ are correspondingly the intensities of transitions polarized parallel ($\mathbf{e} \parallel \mathbf{E}$) and perpendicular ($\mathbf{e} \perp \mathbf{E}$) to the direction of the in-plane electric field.}
\label{pic:OscoftheTR}
\end{figure}

Let us consider transitions between the ground state and the first excited state of the Aharonov-Bohm ring in the limit of weak in-plane electric field, $eER \ll \hbar^2/2M_{e}R^2$.
Away from the points of degeneracy the ground and the first excited states are characterized by a particular value of $m$ and either $P_{if}^{-}$ or $P_{if}^{+}$ given by Eqs.~(\ref{PifP}) and (\ref{PifM}) vanishes. As a result, the angular dependence in Eq.~(\ref{MatrixElementS}) disappears and
the transitions have no linear polarization. The picture changes drastically when $\Phi$ is equal to an integer number of $\Phi_{0}/2$. Then, $P_{if}^{-}=P_{if}^{+}$, and therefore the rate of transitions induced by the radiation polarized parallel to the direction of the in-plane electric field ($\theta=0$) is equal to zero, $T_{if}=T_{\parallel}=0$. Simultaneously, $T_{\perp}$, which is the rate of transitions induced by the light polarized perpendicular to the direction of the in-plane electric field ($\theta=\pi/2$), reaches its maximum possible value. This leads to the strong optical anisotropy of the system. The results of our calculations for the whole range of $\Phi$ are shown in Fig.~\ref{pic:OscoftheTR}.
Very sharp peaks at $\Phi$ equal to an integer number of $\Phi_{0}$ are the result of splitting between the first and second excited states, which were degenerate with energy $\varepsilon_{1}(0)$ in the absence of an external electric field (see Fig.~\ref{pic:ESinpresenceofEF}). This splitting occurs in the second order in $eER$ and the spectacular sharpness of the peaks is due to the very fast change in the electron first and second excited-state wave functions when one moves away from the point of degeneracy (for details, see the Appendix). The optical transitions between the electron ground and second excited states are also linearly polarized, but with $\theta=0$, so that the polarization of these transitions is normal to the polarization of transitions between the electron ground and first excited states. Because these two peaks are very closely separated for $\Phi=0$, the polarization effects are strongly suppressed if the finite linewidth of the radiation is taken into account.

In the case of circularly polarized light the degree of polarization oscillates as well. Inter-level transitions between the ``pure'' states, characterized by the definite angular momentum values differing by one, are either right-hand or left-hand polarized. However, one can easily see that transitions involving the states, which are strongly ``mixed'' when the flux is an integer number of $\Phi_0/2$, have the same probabilities for both circular polarizations. Thus, the magnetic-field-induced optical chirality of quantum rings oscillates with the flux.

The total probabilities of the inter-level transitions indeed depend on the populations of the states involved. However, the discussed oscillations of the degree of polarization do not depend on temperature. This effect allows Aharonov-Bohm rings to be used as room-temperature polarization-sensitive detectors of THz radiation or optical magnetometers.

\section{Discussion and conclusions}
\label{S5}
We have demonstrated that a lateral electric field, which is known to suppress Aharonov-Bohm oscillations in the ground-state energy spectrum of a quantum ring, results in strong oscillations of other physical characteristics
of the system. Namely, the electric-field-induced dipole moment oscillates as a function of the magnetic flux piercing the ring, with pronounced maxima when the flux is equal to an odd number of one-half of the flux quantum.
This effect is caused by lifting the degeneracy of states with different angular momentum by arbitrary small electric fields.
It should be emphasized that the discussed effect is not an artifice of the infinitely-narrow ring model used in
our calculations, but it persists in finite-width rings in a uniform magnetic field.
Indeed, the essential feature required for this effect is the degeneracy of the states with the angular momenta
differing by one at certain magnetic field values, which is known to take place for finite-width rings as well.\cite{PortnoiICPS28}

Future observation of the dipole moment magneto-oscillations would require careful tailoring of the ring
parameters and experiment conditions. For example, the size of the quantum ring should not exceed the electron mean free path but should be large enough so that, for experimentally attainable magnetic fields, the flux through the ring is near the flux quantum. The electric field should not be too large to avoid polarizing the ring strongly in the absence of a magnetic field, but it should be large enough to achieve a splitting between the ground and first excited states
exceeding  $k_{\mathrm{B}}T$.
Our estimates show that all of these conditions can be met in existing quantum ring systems. However, the temperature constraint constitutes the major obstacle for any potential applications outside the low-temperature laboratory.

The temperature restrictions are less essential for another predicted effect -- giant magneto-oscillations of the polarization degree of radiation associated with inter-level transitions in Aharonov-Bohm rings.
Notably, these transitions for the rings satisfying the remaining constraints should occur at THz frequencies. Creating reliable, portable, and tunable sources of THz radiation is one of the most formidable problems of contemporary applied physics. The unique position of the THz range between the frequencies covered by existing electronic or optical mass-produced devices results in an unprecedented variety of ideas aiming to bridge the so-called THz gap; for example, the proposed methods of down-conversion of optical excitations range from creating ultra-fast saturable absorbers \cite{Avrutin88} and utilizing the magnetic-field-induced energy gap in metallic carbon nanotubes\cite{PortnoiSM08,Portnoi09,Batrakov10,CNTexciton} to recent proposals of exciting THz transitions between exciton-polariton branches in semiconductor microcavities.\cite{KKavokin10,delValle11,Shelykh11} Arguably, the use of quantum rings for THz generation and detection has its merits, since their electronic properties can be easily tuned by external fields. The following scheme for using Aharonov-Bohm quantum rings as tuneable THz emitters can be proposed. The inversion of population in semiconductor quantum rings or type II quantum dots can be created by optical excitation across the semiconductor gap. Angular momentum and spin conservation rules do not forbid the creation of an electron in the first excited state as long as the total selection rules for the whole system, consisting of an electron-hole pair and a photon causing this transition, are satisfied. Terahertz radiation will be emitted when the electron undergoes a transition from the excited to the ground state of the ring. As was shown in the previous sections, both the frequency and polarization properties of this transition can be controlled by external magnetic and electric fields.

Other potential applications of the discussed effects are in the burgeoning areas of quantum computing and cryptography. The discussed mixing of the two states, which are degenerate in the absence of electric field, is completely controlled by the angle between the in-plane field and a fixed axis. This brings the potential possibility for creating nanoring-based qubits, which do not require weak spin-orbit coupling between the electric field and electron spin. Arrays of the Aharonov-Bohm rings can also be used for polarization-sensitive single-photon detection, which is
essential for quantum cryptography.

\begin{acknowledgements}
This work was supported by the FP7 Initial Training Network Spin-Optronics (Grant No. FP7-237252)
and
FP7 IRSES projects SPINMET (Grant No. FP7-246784), TerACaN (Grant No. FP7-230778),
and ROBOCON (Grant No. FP7-230832). We are grateful to A.\,V.~Shytov for fruitful discussions and to C.\,A.~Downing
for a critical reading of the manuscript and for helpful suggestions.
\end{acknowledgements}

\newpage

\appendix*
\section{Analytical solutions for small matrices}

In the limit of weak electric field, $\beta=eER/(\hbar^2/M_{e}R^2) \ll 1$, the electron ground, first and second excited states are well-described by the following $3\times3$ system, which is obtained from Eq.~(\ref{Linearsystem}) for $\left | m \right | \le 1$
\begin{equation}
\label{3by3ME}
\begin{pmatrix} \left ( f+1 \right )^{2} & \beta & 0 \\ \beta & f^{2} & \beta \\ 0 & \beta & \left ( f-1 \right )^{2} \end{pmatrix} \begin{pmatrix} c_{+1}^{n} \\ c_{0}^{n} \\ c_{-1}^{n} \end{pmatrix} = \lambda_{n} \begin{pmatrix} c_{+1}^{n} \\ c_{0}^{n} \\ c_{-1}^{n} \end{pmatrix} \mbox{.}
\end{equation}
Here $f=(\Phi-N\Phi_0)/\Phi_0$ with $N$ integer, so that $0 \le f \le 1/2$. The eigenvalues $\lambda_{n}$ of the system (\ref{3by3ME}) are the roots of the cubic equation
\begin{equation}
\label{3Equation}
\lambda_{n}^{3} - \lambda_{n}^{2} \left( 3 f^{2} + 2 \right)  + \lambda_{n} \left( 3 f^{4} + 1 - 2 \beta^{2} \right) -f^{6} +2f^{4} -f^{2} + 2 f^{2} \beta^{2} + 2 \beta^{2} =0 \mbox{.}
\end{equation}
Solving Eq.~(\ref{3Equation}) we find
\begin{equation}
\label{3x3EVa1}
\lambda_{1} = - 2/3 \sqrt{1+ 12 f^{2} +6 \beta^2} \cos \left( \alpha /3 \right) + f^{2}+2/3 \mbox{,}	
\end{equation}
\begin{equation}
\label{3x3EVa2}
\lambda_{2} = - 2/3 \sqrt{1+ 12 f^{2} +6 \beta^2} \cos \left( \alpha /3 - 2 \pi /3 \right ) + f^{2}+2/3 \mbox{,}	
\end{equation}
\begin{equation}
\label{3x3EVa3}
\lambda_{3} = - 2/3 \sqrt{1+ 12 f^{2} +6 \beta^2} \cos \left( \alpha /3 + 2 \pi /3 \right ) + f^{2}+2/3 \mbox{,}	
\end{equation}
with
$$\cos{\alpha}=\frac { 1 - 36 f^2 + 9 \beta^{2}}  { \left (1 + 12 f^{2} +6 \beta^2 \right )^{3/2}} \mbox{.}$$
Considering $\beta \ll 1$ (the limit of weak electric field) we expand Eqs.~(\ref{3x3EVa1}-\ref{3x3EVa3}) into the Taylor series in $f$ to obtain
\begin{equation}
\label{3x3EVa1s}
\lambda_{1} = f^2 - 2 \beta^2 \sum \limits_{n=0}^{\infty} \left ( 2f \right )^{2n} + O(\beta^4) \mbox{, }	
\end{equation}
\begin{equation}
\label{3x3EVa2s}
\lambda_{2} = 1+f^{2}+\beta^{2} \left [ 1- \sum \limits_{n=0}^{\infty} \frac {\left ( -1 \right)^{n} \left ( 2n \right)!} {\left (1-2n \right) \left ( n! \right)^{2}} \left( \frac {f} {\beta^{2}} \right)^{2n} \right ] + O(\beta^4)\mbox{, }	
\end{equation}
\begin{equation}
\label{3x3EVa3s}
\lambda_{3} = 1+f^{2}+\beta^{2} \left [ 1+ \sum \limits_{n=0}^{\infty} \frac {\left ( -1 \right)^{n} \left ( 2n \right)!} {\left (1-2n \right) \left ( n! \right)^{2}} \left( \frac {f} {\beta^{2}} \right)^{2n} \right ] + O(\beta^4)\mbox{. }	
\end{equation}
It can be shown that Eqs.~(\ref{3x3EVa2s}) and (\ref{3x3EVa3s}) coincide with the results of the perturbation theory in $eER$ for quasi-degenerate states\cite{Messiah} if the coupling to the states with $|m|>1$ is neglected.

\begin{figure}[t]%
\centering
\includegraphics*[width=0.7\linewidth]{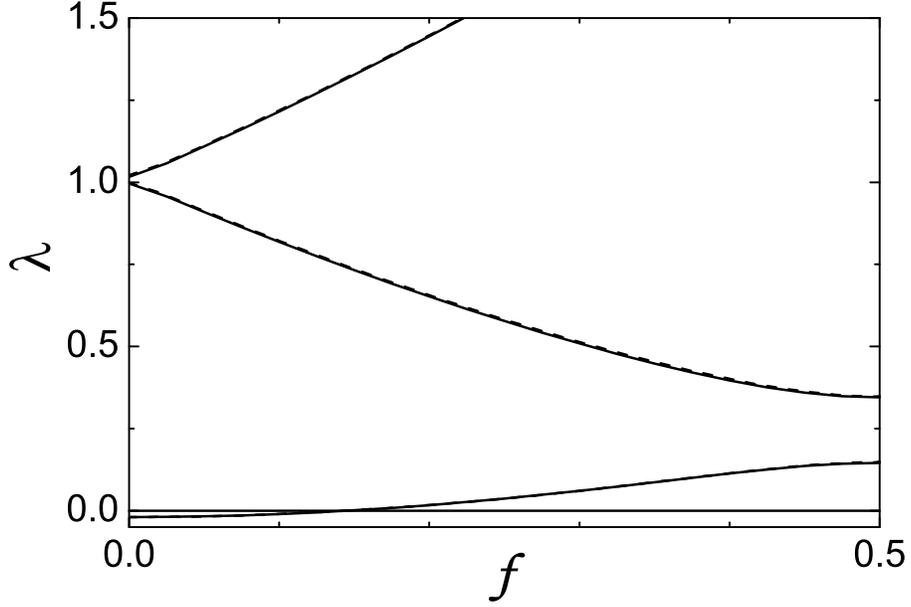}
\caption{The normalized energy spectrum as a function of dimensionless parameter $f$ for $\beta=0.1$. Dashed line: the result of analytical solution of the $3 \times 3$ system. Solid line: the result of numerical diagonalization of the $23 \times 23$ system. A horizontal line is shown to indicate $\lambda=0$ value.}
\label{pic:ES3x3}
\end{figure}

 The energy spectrum given by Eqs.~(\ref{3x3EVa1}--\ref{3x3EVa3}) is plotted in Fig.~\ref{pic:ES3x3}. It is nearly indistinguishable from the energy spectrum, which was obtained by numerical diagonalization of the $23 \times 23$ system in Sec.~\ref{S2} for the same value of $\beta$.
A small discrepancy between the plotted energy spectra is noticeable only for the first and second excited states. The energy spectrum obtained by numerical diagonalization of the $23 \times 23$ system is slightly shifted towards the smaller energies. This shift occurs because the considered $3 \times 3$ matrix does not take into account the coupling between the $m=\pm 1$ and $m=\pm 2$ states. For the infinite system and $f=0$, perturbation theory up to the second order in $\beta$ yields
\begin{equation}
\label{Lambdas03by3}
\lambda_{1}=-2 \beta^{2} \mbox{, } \lambda_{2}=1 - \beta^{2}/3\mbox{, } \lambda_{3}=1 + 5\beta^{2}/3\mbox{, }
\end{equation}
whereas from Eqs.~(\ref{3x3EVa1s}-\ref{3x3EVa3s}) one gets
\begin{equation}
\label{Lambdas0PT}
\lambda_{1}=-2 \beta^{2} \mbox{, } \lambda_{2}=1 \mbox{, } \lambda_{3}=1 + 2\beta^{2} \mbox{. }
\end{equation}
The $\lambda_{2}$ and $\lambda_{3}$ values in Eq.~(\ref{Lambdas03by3}) differ from the values in Eq.~(\ref{Lambdas0PT}) by $-\beta^{2}/3$ which corresponds to the repulsion between the $m=\pm1$ and $m=\pm2$ states calculated using the second-order perturbation theory.

When $f=1/2$, and in the absence of a lateral electric field, the $m=0$ and $m=-1$ states are degenerate with energy $\varepsilon_{1} \left ( 0 \right)/4$, i.e. $\lambda_1=\lambda_2=1/4$, whereas the $m=+1$ state energy is nine times larger ($\lambda_3=9/4$). The contribution from this remote state can be neglected, and the electron ground and first excited states are well-described by the following $2\times2$ system, which contains $c_{-1}$ and $c_{0}$ coefficients only:
\begin{equation}
\label{2by2ME}
\begin{pmatrix} f^{2} & \beta \\ \beta & \left ( f-1 \right )^{2} \end{pmatrix} \begin{pmatrix} c_{0}^{n} \\ c_{-1}^{n} \end{pmatrix} = \lambda_{n} \begin{pmatrix} c_{0}^{n} \\ c_{-1}^{n} \end{pmatrix} \mbox{.}
\end{equation}
\begin{figure}[t]%
\centering
\includegraphics*[width=0.7\linewidth]{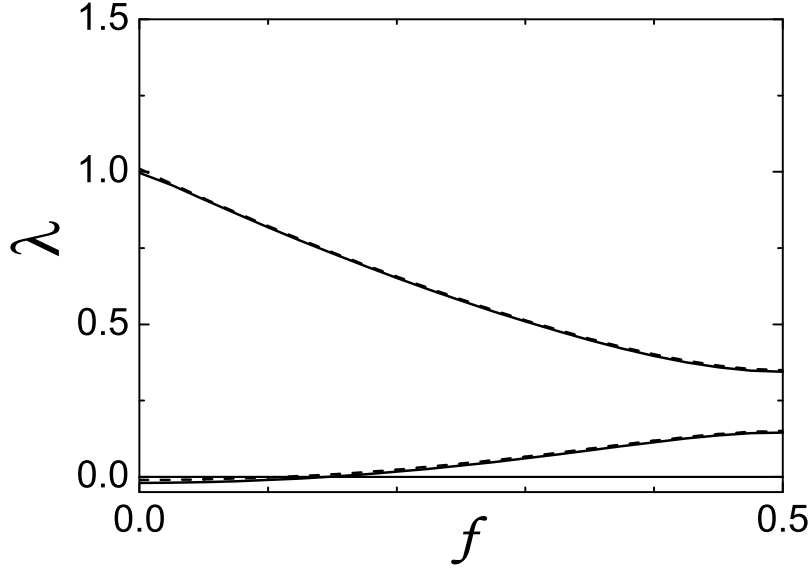}
\caption{The normalized energy spectrum as a function of dimensionless parameter $f$ for $\beta=0.1$. Dashed line - the result of analytical solution of the $2 \times 2$ system. Solid line - the result of numerical diagonalization of the $23 \times 23$ system. A horizontal line is shown to indicate $\lambda=0$ value.}
\label{pic:ES2x2}
\end{figure}
The eigenvalues $\lambda_{n}$ of the system (\ref{2by2ME}) are the roots of the quadratic equation
\begin{equation}
\label{2Equation}
\lambda_{n}^2 -\lambda_{n} \left( 2 f^2 - 2f +1 \right) +f^{4} -2f^{3} +f^{2} -\beta^2 =0 \mbox{.}
\end{equation}
Solving Eq.~(\ref{2Equation}) we find
\begin{equation}
\label{2x2EVa}
\lambda_{1,2}= f^{2}-f+1/2 \mp \sqrt{f^{2}-f + \beta^{2} + 1/4} \mbox{,}
\end{equation}
yielding for $f=1/2$ the eigenvalue difference $\lambda_2-\lambda_1=2\beta$, corresponding to the energy splitting of $eER$ as expected from the perturbation theory for degenerate states. The energy spectrum given by Eq.~(\ref{2x2EVa}) is plotted in Fig.~\ref{pic:ES2x2} together with two lowest eigenvalues of the $23 \times 23$ system demonstrating a spectacular accuracy of the approximate solution for $\beta=0.1$.

Let us now return to the $3\times3$ matrix and examine how its eigenvectors are modified with changing $f$. 
Near the point $f=0$, it is convenient to write the eigenvectors of the system (\ref{3by3ME}) in the following form:
\begin{equation}
\label{3x3EVe}
\begin{pmatrix} c_{+1}^{n} \\ c_{0}^{n} \\ c_{-1}^{n} \end{pmatrix} = A_{n} \begin{pmatrix} \left [ \lambda_{n}  - \left( f -1 \right)^{2} \right] \left ( \lambda_{n} -f^2 \right ) - \beta^{2} \\ \left [ \lambda_{n} - \left( f-1 \right)^{2} \right]  \beta \\ \beta^{2} \end{pmatrix} \mbox{,}
\end{equation}
where $A_{n}$ denotes the normalization constant corresponding to the eigenvalue $\lambda_n$, and (\ref{3x3EVe}) is valid only for $\beta \ne 0$.
For $f=0$ in the limit of weak electric field ($\beta \ll 1$), we obtain
\begin{equation}
\label{3x3EVe1f0}
\begin{pmatrix} c_{+1}^{1} \\ c_{0}^{1} \\ c_{-1}^{1} \end{pmatrix} = \frac{ \left ( 1+ 1 \sqrt{1+8 \beta^{2}} + 8 \beta^{2} \right )^{-1/2}} {\sqrt{2}} \begin{pmatrix} - 2 \beta \\ 1 + \sqrt{1+8 \beta^{2}} \\ - 2 \beta \end{pmatrix}  \stackrel{\beta \to 0} {\longrightarrow} \begin{pmatrix} 0 \\ 1  \\ 0 \end{pmatrix} \mbox{,}
\end{equation}
\begin{equation}
\label{3x3EVe2f0}
\begin{pmatrix} c_{+1}^{2} \\ c_{0}^{2} \\ c_{-1}^{2} \end{pmatrix} = \frac {1} {\sqrt{2}} \begin{pmatrix} - 1 \\ 0 \\ 1 \end{pmatrix} \mbox{, }
\end{equation}
\begin{equation}
\label{3x3EVe3f0}
\begin{pmatrix} c_{+1}^{3} \\ c_{0}^{3} \\ c_{-1}^{3} \end{pmatrix} =  \frac{ \left ( 1- 1 \sqrt{1+8 \beta^{2}} + 8 \beta^{2} \right )^{-1/2}} {\sqrt{2}} \begin{pmatrix} 2 \beta \\ \sqrt{1+8 \beta^{2}} - 1 \\ 2 \beta \end{pmatrix} \stackrel{\beta \to 0} {\longrightarrow} \frac {1} {\sqrt{2}} \begin{pmatrix} 1 \\ 0 \\ 1 \end{pmatrix} \mbox{.}
\end{equation}
From Eqs.~(\ref{3x3EVe1f0}--\ref{3x3EVe3f0}), one can see that for $f=0$ and $\beta \ll 1$, the electron ground state is almost a pure $m=0$ state, whereas the angular dependencies of the wavefunctions of the first and second excited states are well-described by $\sin \varphi$ and $\cos \varphi$, respectively.

The structure of eigenfunctions near $f=1/2$ is best understood from Eq.~(\ref{2by2ME}), which yields
\begin{equation}
\label{2x2EVe1}
\begin{pmatrix} c_{0}^{1} \\ c_{-1}^{1} \end{pmatrix} = A \begin{pmatrix} \beta  \\ 1/2 - f - \sqrt{f^{2}-f + \beta^{2} + 1/4}   \end{pmatrix} \mbox{,}
\end{equation}
\begin{equation}
\label{2x2EVe2}
\begin{pmatrix} c_{0}^{2} \\ c_{-1}^{2} \end{pmatrix} = A \begin{pmatrix} f - 1/2 + \sqrt{f^{2}-f+ \beta^{2} + 1/4 } \\ \beta \end{pmatrix} \mbox{.}
\end{equation}
Here $A$ is the normalization constant and $\beta \ne 0$. For $f=1/2$ we obtain
\begin{equation}
\label{2x2EVef05}
\begin{pmatrix} c_{0}^{1} \\ c_{-1}^{1} \end{pmatrix} = \frac {1} {\sqrt{2}} \begin{pmatrix} 1 \\ - 1 \end{pmatrix} \mbox{, } \begin{pmatrix} c_{0}^{2} \\ c_{-1}^{2} \end{pmatrix} = \frac {1} {\sqrt{2}} \begin{pmatrix} 1 \\ 1 \end{pmatrix} \mbox{.}
\end{equation}
From Eq.~(\ref{2x2EVef05}) one can see that for $f=1/2$ the angular dependencies of the ground and first excited states wavefunctions are described by $\sin \left ( \varphi/2 \right )$ and $\cos \left ( \varphi/2 \right )$ respectively.

\begin{figure}[t]
\includegraphics*[width=0.47\textwidth]{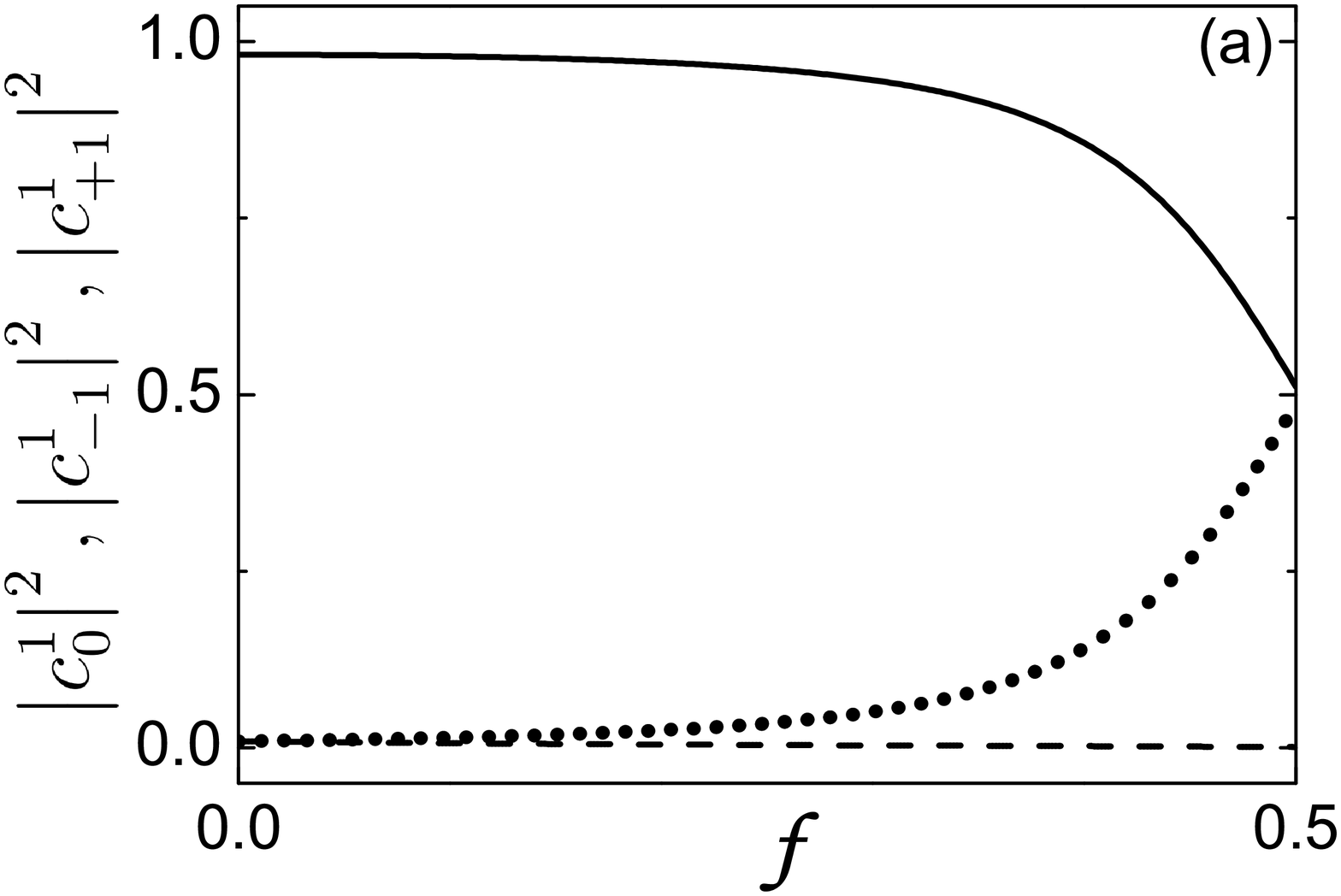}
\includegraphics*[width=0.47\textwidth]{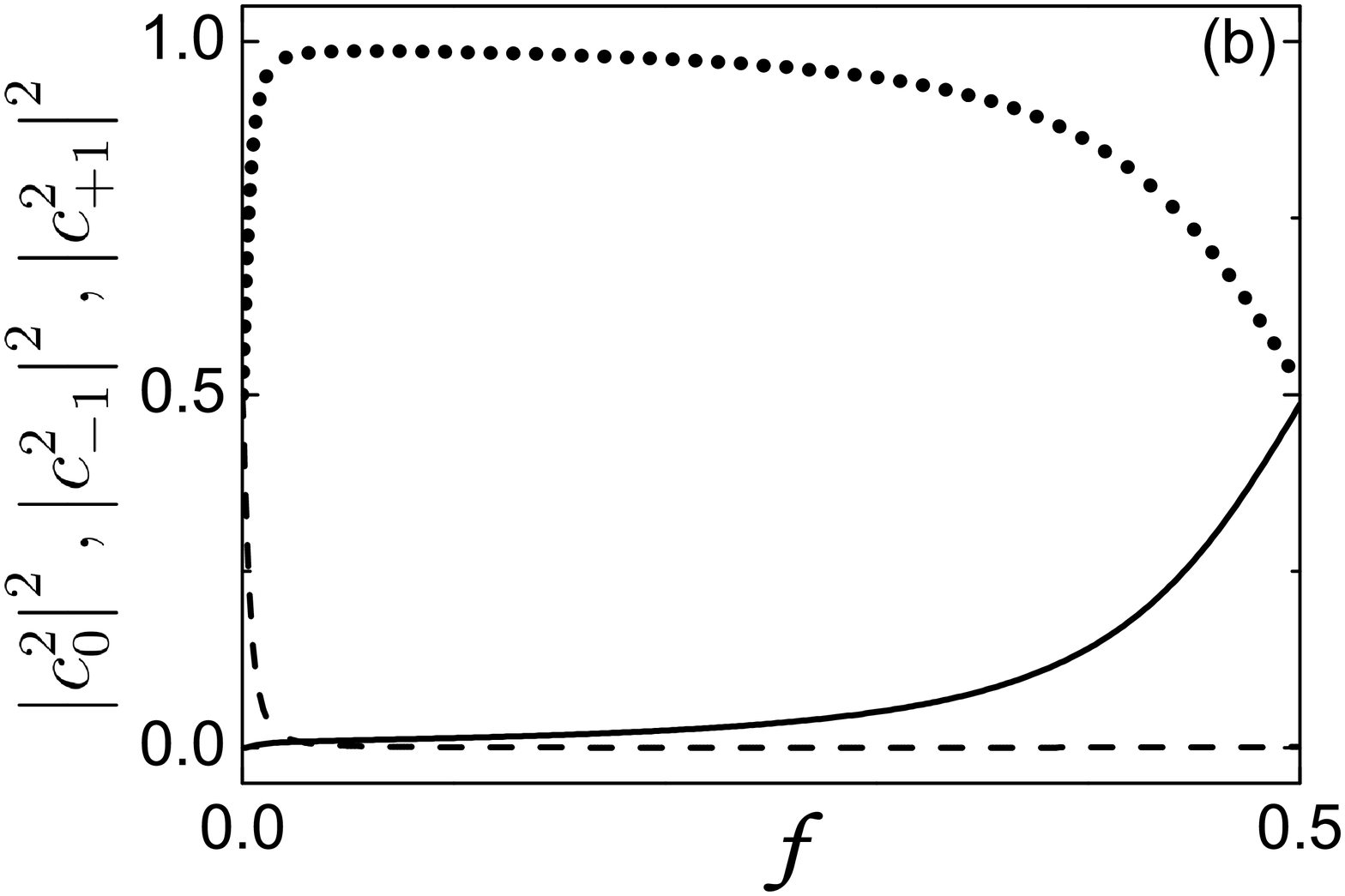}
\includegraphics*[width=0.47\textwidth]{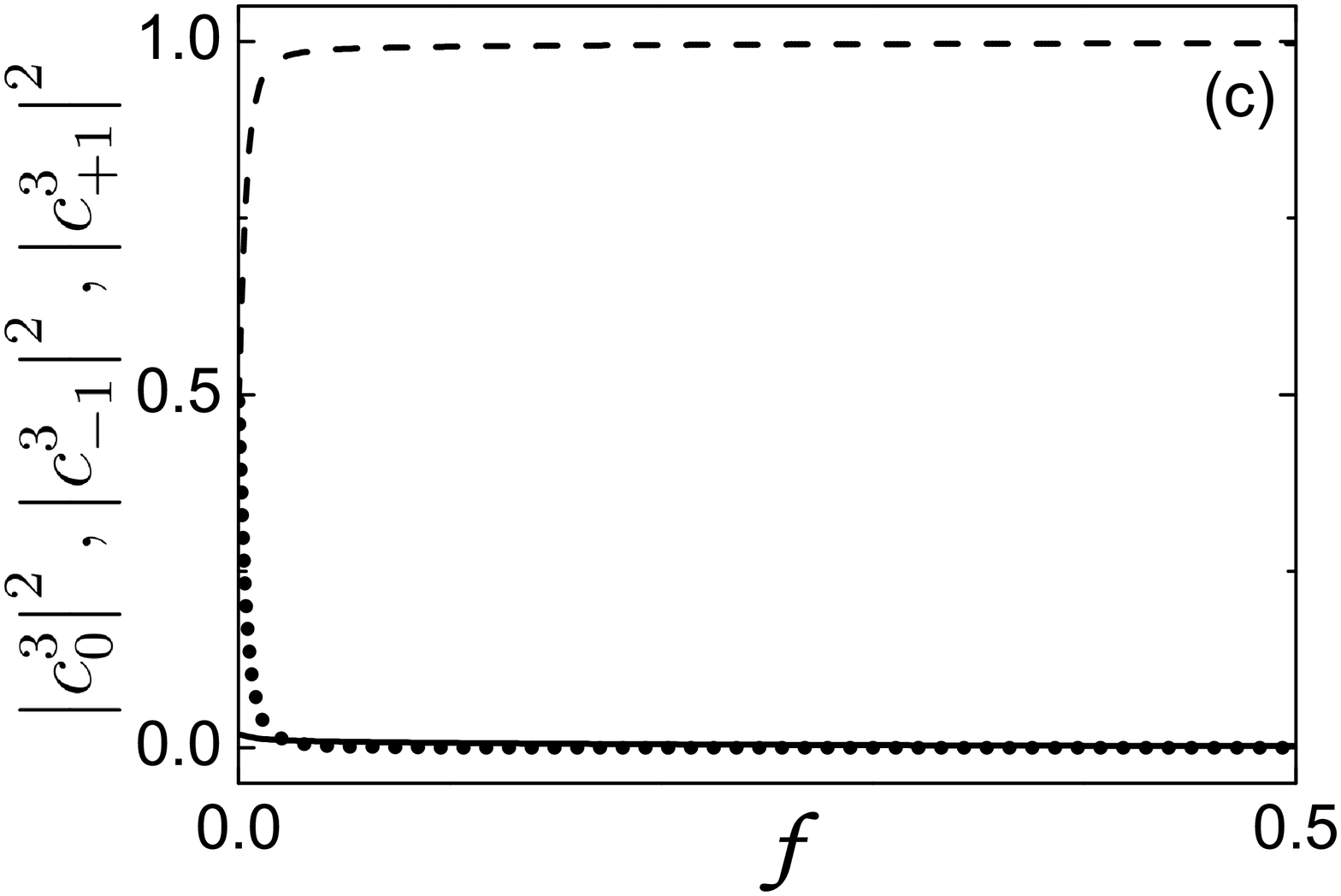}
\caption{Magnetic flux dependence of the wavefunction coefficients $\left | c_{0} \right |^{2}$ (solid line), $\left | c_{-1} \right |^{2}$ (dotted line), and $\left | c_{+1} \right |^{2}$ (dashed line): (a) for the ground state, (b) for the first excited state, and (c) for the second excited state.}
\label{pic:WFC}
\end{figure}

Figure~\ref{pic:WFC} shows the magnetic-flux dependencies of the coefficients  $\left | c_{0} \right |^{2}$, $\left | c_{-1} \right |^{2}$, and $\left | c_{+1} \right|^{2}$ for the electron ground, first, and second excited states.
From these plots, one can see that the electron ground state is almost a pure $m=0$ state in a wide region $0 \le f \lesssim 1/4$. An admixture of the $m=-1$ wavefunction increases smoothly as we approach the point of degeneracy $f=1/2$. Finally, when $f=1/2$, the ground-state wave function is expressed as a difference of the $m=-1$ and $m=0$ wave functions. The first and second excited states behave differently. In a small region near the point $f=0$, the electron first and second excited-state wave functions consist of a strong mixture of the $m=-1$ and $m=+1$ functions with a tiny admixture of the $m=0$ function. In particular, when $f=0$, the first and second excited-state eigenfunctions with good accuracy can be expressed as a difference and a sum of the $m=-1$ and $m=+1$ functions respectively. Optical transitions between these states and the ground state are only allowed if the polarization of the associated optical excitations is either perpendicular (for the first excited state) or parallel (for the second excited state) to the direction of the applied in-plane electric field. Away from the $f=0$ region, only the coefficient $c_{-1}$ (in the case of the first excited state) or $c_{+1}$ (in the case of the second excited state) remains in Eq.~(\ref{3x3EVe}), which now describes almost pure $m=+1$ and $m=-1$ states. When $f$ exceeds $1/4$, the first excited state starts to contain a noticeable ad-mixture of $m=0$ function, as discussed above, and for $f=1/2$ the first excited-state eigenfunction is expressed as a sum of the $m=-1$ and $m=0$ wave functions in equal proportions, whereas the second excited state remains an almost pure $m=+1$ state.

The same trend in the evolution of wave functions of the three lowest-energy states with changing the flux through the ring can be seen from perturbation theory. For $f=0$, the degeneracy between the first and second excited states is removed in the second order in $eER$ only. Nevertheless, as a result of the degeneracy, the introduction of any weak perturbation drastically modifies the wavefunctions corresponding to these states, turning them from the eigenstates of the angular momentum operator to the sine and cosine functions. With a slight increase of $f$, so that $f>\beta^2$, the first and the second excited states, which are not degenerate anymore for $f \neq 0$, become governed mainly by the diagonal terms of the Hamiltonian, which do not mix the $m=-1$ and $m=+1$ functions. When $f=1/2$, the $m=-1$ and $m=0$ states are degenerate in the absence of the electric field. This degeneracy is removed in the first order in $eER$. The off-diagonal matrix elements connecting the $m=-1$ and $m=0$ functions remain of the same order of magnitude as the difference between the diagonal terms of the Hamiltonian across a broad range of $f$ values near $f=1/2$. This results in strong mixing of the $m=-1$ and $m=0$ components in the eigenfunctions of the ground and first excited states for $1/4 \lesssim f \le 1/2$.


\newpage

\begin{thebibliography}{[100]}

\bibitem{Lorke}%
 A. Lorke, R.\,J. Luyken, A.\,O. Govorov, J.\,P. Kotthaus, J.\,M. Garcia, and P.\,M. Petroff,
 Phys. Rev. Lett. \textbf{84}, 2223 (2000).

\bibitem{Ribeiro}%
 E. Ribeiro, A.\,O. Govorov, W. Carvalho, Jr., and G. Medeiros-Ribeiro,
 Phys. Rev. Lett. \textbf{92}, 126402 (2004).

\bibitem{Chen}%
 J.~Chen, W.\,S.~Liao, X.~Chen, T.~Yang, S.\,E. Wark, D.\,H. Son, J.\,D. Batteas, and P.\,S. Cremer,
 ACS Nano \textbf{3}, 173 (2009).

\bibitem{review1}%
 A.\,G. Aronov and Yu.\,V. Sharvin,
 Rev. Mod. Phys. \textbf{59}, 755 (1987).

\bibitem{review2}%
 S. Viefers, P. Koskinen, P. Singha~Deo, and M. Manninen,
 Physica E (Amsterdam) \textbf{21}, 1 (2004).

\bibitem{review3}%
 T. Ihn, A. Fuhrer, L. Meier, M. Sigrist, and K. Ensslin,
 Europhysics News \textbf{36}, 78 (2005).

\bibitem{Siday}%
 W. Ehrenberg and R.\,E. Siday,
 Proc. Phys. Soc. London Sect. B \textbf{62}, 8 (1949).

\bibitem{Aharonov-Bohm}%
 Y. Aharonov and D. Bohm,
 Phys. Rev. \textbf{115}, 485 (1959).

\bibitem{Barticevic}%
 Z. Barticevic, G. Fuster, and M. Pacheco,
 Phys. Rev. B \textbf{65}, 193307 (2002).

\bibitem{Bruno-Alfonso}%
 A. Bruno-Alfonso and A. Latg\'e,
 Phys. Rev. B, \textbf{71}, 125312 (2005).

\bibitem{Fischer}%
 A.\,M. Fischer, V.\,L. Campo, Jr., M.\,E. Portnoi, and R.\,A. R\"{o}mer,
 Phys. Rev. Lett. \textbf{102}, 096405 (2009).

\bibitem{Vivaldo}%
 M.\,D. Teodoro, V.\,L. Campo, Jr., V. Lopez-Richard, E. Marega, Jr., G.\,E. Marques, Y. Galv\~{a}o~Gobato, F. Iikawa, M.\,J.\,S.\,P. Brasil,
 Z.\,Y. AbuWaar, V.\,G. Dorogan, Yu.\,I. Mazur, M. Benamara, and G.\,J. Salamo
 Phys. Rev. Lett. \textbf{104}, 086401 (2010).

\bibitem{helix1}%
 O.\,V. Kibis, S.\,V. Malevannyy, L. Huggett, D.\,G.\,W. Parfitt, M.\,E. Portnoi,
 Electromagnetics \textbf{25}, 425 (2005).

\bibitem{helix2}%
 O.\,V. Kibis and M.\,E. Portnoi,
 Tech. Phys. Lett. \textbf{33}, 878 (2007).

 \bibitem{helix3}%
  O.\,V. Kibis and M.\,E. Portnoi,
  Physica E \textbf{40}, 1899 (2008).

\bibitem{PortnoiICPS28}%
 M.\,E. Portnoi, O.\,V. Kibis, V.\,L. Campo~Jr, M. Rosenau da Costa, L. Huggett, and S.\,V. Malevannyy,
 Proc. 28th ICPS, AIP Conference Proceedings \textbf{893}, 703 (2007).

\bibitem{Avrutin88}%
 E.\,A. Avrutin and M.\,E. Portnoi, Fiz. Tekh. Poluprovodn. \textbf{22}, 1524 (1988)
 [Sov. Phys. Semicond. \textbf{22}, 968 (1988)].

\bibitem{PortnoiSM08}%
 M.\,E. Portnoi, O.\,V. Kibis, M. Rosenau da Costa,
 Superlatt. Microstruct. \textbf{43}, 399 (2008).

\bibitem{Portnoi09}%
 M.\,E. Portnoi, M. Rosenau da Costa, O.\,V. Kibis, and I.\,A. Shelykh,
  Int. J. Mod. Phys. B \textbf{23}, 2846 (2009).

\bibitem{Batrakov10}%
 K.\,G. Batrakov, O.\,V. Kibis, P.\,P. Kuzhir, M. Rosenau da Costa, and M.\,E. Portnoi, J. Nanophoton. \textbf{4}, 041665 (2010).

 \bibitem{CNTexciton}%
 R.\,R. Hartmann, I.\,A. Shelykh, and M.\,E. Portnoi, Phys. Rev. B \textbf{84}, 035437 (2011).

\bibitem{KKavokin10}%
 K.\,V. Kavokin, M.\,A. Kaliteevski, R.\,A. Abram, A.\,V. Kavokin, S. Sharkova, and I.\,A. Shelykh,
 Appl. Phys. Lett. \textbf{97}, 201111 (2010).

\bibitem{delValle11}%
 E.~del~Valle and A. Kavokin,
 Phys. Rev. B \textbf{83}, 193303 (2011).

\bibitem{Shelykh11} I.\,G. Savenko, I.\,A. Shelykh, and M.\,A. Kaliteevski, Phys. Rev. Lett. \textbf{107}, 027401 (2011).

\bibitem{Messiah} A. Messiah, \textit{Quantum Mechanics} (North-Holland, Amsterdam, 1965), Vol. 2, p. 711.

\end{thebibliography}
\end{document}